\documentclass[twocolumn,prl,english,reprint, longbibliography, superscriptaddress, breaklinks=true, showkeys, showpacs=false, nofootinbib]{revtex4}
\usepackage{color,footnote}
\usepackage{babel}
\usepackage{amsmath,amssymb}
\usepackage{mathtools,bbm}
\usepackage{subfigure}
\usepackage{graphicx}
\usepackage{physics,braket}
\usepackage{comment}
\usepackage{mathtools}
\usepackage[varg]{txfonts} 
\usepackage[unicode=true,pdfusetitle,bookmarks=true,bookmarksnumbered=false,bookmarksopen=false,breaklinks=true,pdfborder={0 0 0},backref=false,colorlinks=true]
 {hyperref}
\makeatletter
\@ifundefined{textcolor}{}
{%
 \definecolor{BLACK}{gray}{0}
 \definecolor{WHITE}{gray}{1}
 \definecolor{RED}{rgb}{1,0,0}
 \definecolor{GREEN}{rgb}{0,1,0}
 \definecolor{BLUE}{rgb}{0,0,1}
 \definecolor{CYAN}{cmyk}{1,0,0,0}
 \definecolor{MAGENTA}{cmyk}{0,1,0,0}
 \definecolor{YELLOW}{cmyk}{0,0,1,0}
}
\pdfoutput=1
\hypersetup{colorlinks=true,citecolor=blue,linkcolor=cyan,urlcolor=blue,filecolor= green, breaklinks=true}
\usepackage{url}
\usepackage{breakurl}
\makeatother

\newcommand{\mf}{\mathfrak}

\begin{document}

\author{D. S. Starke\href{https://orcid.org/0000-0002-6074-4488} {\includegraphics[scale=0.05]{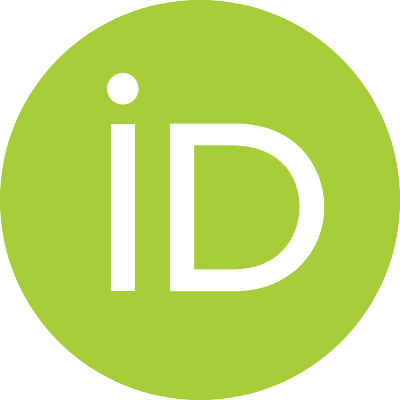}}}
\email{starkediego@gmail.com (corresponding author)}
\affiliation{Physics Department, 
Federal University of Santa Maria, 97105-900,
Santa Maria, RS, Brazil}

\author{J. Maziero\href{https://orcid.org/0000-0002-2872-986X}{\includegraphics[scale=0.05]{orcidid.pdf}}}
\email{jonas.maziero@ufsm.br}
\affiliation{Physics Department, 
Federal University of Santa Maria, 97105-900,
Santa Maria, RS, Brazil}

\author{R. M. Angelo\href{https://orcid.org/0000-0002-7832-9821}{\includegraphics[scale=0.05]{orcidid.pdf}}}
\email{renato.angelo@ufpr.br}
\affiliation{Department of Physics, Federal University of Paran\'a, Curitiba, Paran\'a, P.O. Box 19044, 81531-980, Brazil}

\selectlanguage{english}


\title{
Correlating Local Quantum Reality with Causally Disconnected Choices
}

\begin{abstract}
In 1935, Einstein, Podolsky, and Rosen (EPR) claimed the incompleteness of quantum mechanics based on the notions of realism (``{\it If, without in any way disrupting a system, we can predict with certainty - i.e., with a probability of one - the value of a physical quantity, then an element of physical reality corresponds to this physical quantity.}'')  and locality (``{\it ...\,since the two systems no longer interact, no real change can take place in the second system in consequence of anything that may be done to the first system}''). EPR also insisted that ``{\it The elements of physical reality cannot be determined by \emph{a priori} philosophical considerations, but must be found by\,...\,experiments and measurements.}''. Here, employing an operational framework for testing realism in quantum systems, we envisage an experiment -- referred to as the Reality Quantum Correlator (RQC) -- capable of showing that the elements of reality in one laboratory can be correlated with causally disconnected choices, thus questioning EPR's locality. Empirical evidence supporting our theoretical predictions is then provided by implementing the corresponding quantum circuit on IBM's quantum computers.
\end{abstract}



\date{\today}

\maketitle

\section{Introduction}

Quantum mechanics, mainly developed around 1925 by Heisenberg, Schr\"odinger, Dirac, Pauli, Jordan, and Born~\cite{Heisenberg1925, Born1925, Dirac1925, Pauli1925, BornH1926, Born1926, Schrodinger1926, Born1927}, prompted the scientific community to mark this year in honor of the 100th anniversary of the theory.
One of the significant debates around the theory was influenced by pre-1900s physical theories, which were fundamentally rooted in two taken-for-granted tenets: realism and local causality. Realism assumes that the physical properties of all systems are well defined regardless of any observers' interventions. Local causality, on the other hand, suggests that effects can only be derived from nearby causes. From these premises, it readily follows that physical reality can by no means be influenced at a distance. Quantum mechanics, it turns out, seems to stand in direct opposition to this state of affairs. This discrepancy prompted Einstein, Podolsky, and Rosen (EPR) to boldly challenge the theory. Unwilling to negotiate the assumption of local causality, the authors put forward a rationale claiming that quantum mechanics is an incomplete model of the physical world~\cite{EPR1935}.

It is widely accepted nowadays that Bell inequality violations proved EPR wrong. The point, nevertheless, deserves circumspection. As Bell showed~\cite{Bell1964,Bell2001} and numerous sophisticated experiments confirmed~\cite{Hensen2015,Giustina2015,Shalm2015,Hensen2016,Rauch2018,Li2018}, a Bell inequality violation implies that Nature is not compatible with the hypothesis of local causality. However, intense debate~\cite{Khrennikov2009,Tresser2010,Gisin2012,Maudlin2014,Zukowski2014,Wiseman2014,Wiseman2015,Gillis2015,Norsen2015,Santos2016,Wiseman2017,Abellan2018,Khrennikov2020,Cavalcanti2021} culminated with local causality being acknowledged as a set of hypotheses, often incorporating the assumptions of realistic hidden variables, freedom of choice and locality. Although many researchers argue that Bell tests disprove locality, thus leading to the phenomenon known as Bell nonlocality~\cite{Brunner2014}, the fact remains that such tests do not definitively resolve the question of which of these hypotheses, if not all of them, is in conflict with Nature's actual behavior.

In this letter, we discuss an experiment which ultimately subscribes to the quantum nonlocality perspective.
We raise and answer the following question: Can the elements of reality in Bob's laboratory be correlated with causally disconnected choices made in Alice's laboratory? We address this question within the framework of an optical setup that can be feasibly implemented using current technologies \cite{Ma2013}. Furthermore, we provide demonstrative evidence (modulo a locality loophole) to support our claims through quantum state tomography experiments conducted on IBM's quantum computers~\cite{IBMQ}.

A great deal of work has been made regarding quantum nonlocality \cite{Brunner2014} and quantum steering \cite{Uola2020}. But while in these works realism is considered through hidden variables or hidden states, here we use an operational approach that was developed \cite{Bilobran2015} to extend the original EPR's definition \cite{EPR1935}: \textit{``A sufficient condition for the reality of a physical quantity is the possibility of predicting it with certainty, without disturbing the system.''}
Much attention has been given also to the quantum eraser \cite{Ma2016}, but our article also goes well beyond these developments by introducing a procedure to quantify the wave-like behavior without the need for retrodiction, thus avoiding the associated paradoxes.

\begin{figure*}[ht]
    \centering
    \includegraphics[scale=1.0]{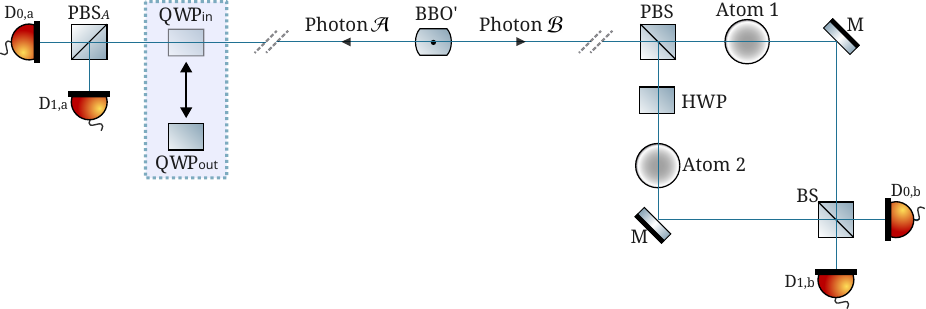}
    \caption{Schematic representation of the reality quantum correlator (RQC), an optical setup with atoms 1 and 2 respectively placed in the upper and lower paths of a Mach-Zehnder interferometer (MZI). 
    In this setting, QWP stands for a quarter wave-plate, PBS (as PBS$_{A}$) is a polarizing beam splitter, HWP is a half wave plate, M represents mirrors, BS is a beam splitter, $D_{j,k}$ are detectors (with $j=0,1$ and $k=a,b$, where $a$ ($b$) is the path of the photon $\mathcal{A}$ ($\mathcal{B}$)), and BBO$^{\prime}$ stands for a non-linear crystal of beta barium borate supplemented with auxiliary equipment, through which a pair of photons is produced with their polarizations partially entangled.
    The $\cal{A}$ and $\cal{B}$ photons go to spacelike separated regions so that they cannot communicate during the time interval needed for the local operations.
    Photon $\mathcal{A}$ can pass through a QWP and goes through the PBS$_{A}$ while the photon $\mathcal{B}$ goes through an MZI. 
    The dotted blue box represents the two arrangements adopted for implementing the reality correlator: for the QWP$_\text{out}$ (QWP$_\text{in}$) configuration, QWP is out (in) of photon $\cal{A}$'s path and a procedure exists that nonlocally correlates Alice's choices with photon $\cal{B}$'s path and atoms' energies being (not being) elements of the physical reality.
    }
    \label{fig:RQC}
\end{figure*}

The subsequent sections of this article are organized in the following manner. In Sec.~\ref{sec:theory}, we introduce the quantum reality quantifier, which we utilize to analyze the reality quantum correlator.
In Sec.~\ref{sec:exp}, we present our experimental proposal for a modified quantum eraser experiment utilizing entangled photons and investigate the incorporation of extra non-interacting degrees of freedom. In Sec.~\ref{sec:analysis}, we provide theoretical predictions and establish the conditions necessary for the erasure of quantum reality at Bob's laboratory. Section~\ref{sec:QC} presents the implementation of the experimental configuration on IBM's quantum computers, followed by the final remarks in Sec.~\ref{sec:FR}.

\section{Quantum Reality: Theoretical framework}
\label{sec:theory}

The key to our argument will be a measure of quantum irrealism that has recently received much attention, from both the theoretical~\cite{Rudnicki2018,Dieguez2018,Freire2019,Lustosa2020,Savi2021,Gomes2022,Orthey2022,Basso2022,Paiva2023,Engelbert2023,Orthey2025,Orthey2025_2} and the experimental sides~\cite{Mancino2018,Dieguez2022}.
From a broader perspective, quantum irrealism is also related to additional foundational elements \cite{Reid2017,Moreira2019,Zhu2021,Thenabadu2022,Starke2024}.
The idea was introduced by Bilobran and Angelo \cite{Bilobran2015}. From the premise that a measurement of an observable $Q=\sum_iq_iM_i^q$ acting on the Hilbert space $\mathcal{H_Q}$, for a given preparation $\rho$ on $\mathcal{H=H_Q\otimes H_R}$ (with $\mathcal{H_R}$ possibly multipartite), determines an element of reality (a well defined value for $Q$), the authors suggested $\Phi_{Q}(\rho)=\rho$ as a criterion for $Q$'s realism, where $\Phi_Q(\rho)\coloneqq \sum_i\tilde{M}_i^q\rho\tilde{M}_i^q$ denotes a nonselective measurement of $Q$ (a dephasing map in $Q$'s eigenbasis) and $\tilde{M}_i^q=M_i^q\otimes \mathbbm{1}_\mathcal{R}$. A quantum state satisfying $\rho=\Phi_{Q}(\rho)$ is then called an $Q$-reality state, as a $Q$ measurement is not capable of inducing any alteration in the preparation. Indeed, for such a state one has $\Phi_Q\circ\Phi_Q(\rho)=\rho$. To quantify how far the observable $Q$ is from being an element of physical reality, we employ the so-called {\it irreality} measure \cite{Bilobran2015}:
\begin{equation}
\mathfrak{I}_Q(\rho)=S\big(\Phi_Q(\rho)\big)-S(\rho) \label{eq:irr},
\end{equation}
where $S(\rho)=-\Tr(\rho \log_2 \rho)$ is the von Neumann entropy. The irreality of any physical quantity $Q$ is never negative and vanishes if and only if $\rho=\Phi_Q(\rho)$. Moreover, the measure of irreality can be decomposed as $\mathfrak{I}_Q(\rho)=C_Q(\rho_\mathcal{Q})+D_Q(\rho)$, where $C_Q(\rho_\mathcal{Q})=S\big(\Phi_Q(\rho_\mathcal{Q})\big)-S(\rho_\mathcal{Q})$ is the relative entropy of coherence~\cite{Baumgratz2014}, with $\rho_{\mathcal{Q}}=\Tr_{\mathcal{R}}(\rho)$, and $D_Q(\rho)=I_\mathcal{Q:R}(\rho)-I_\mathcal{Q:R}\big(\Phi_Q(\rho)\big)$ represents the discord of measurement $Q$, with $I_\mathcal{Q:R}(\rho)=S(\rho_\mathcal{Q})+S(\rho_\mathcal{R})-S(\rho)$ being the mutual information of $\rho$. By the definition of (one-way) quantum discord, $\mathcal{D_Q}(\rho)\coloneqq \min_Q D_Q(\rho)$~\cite{Ollivier2001,Henderson2001}, one has $\mathcal{D_Q}(\rho)\leq D_Q(\rho)$, which renders $\mathfrak{I}_Q(\rho)\geq C_Q(\rho_\mathcal{Q})+\mathcal{D_Q}(\rho)$. Interestingly, this shows that $Q$ realism is prevented by two forms of quantumness, namely, quantum coherence \cite{Baumgratz2014} and quantum correlations \cite{Modi2012}.

\section{Experimental proposal}
\label{sec:exp}

Our proposal consists of experimentally verifying that the quantum reality of an observable in Bob's laboratory can be correlated with causally disconnected choices in Alice's laboratory. To this end, we conceive the optical setting depicted in Fig.~\ref{fig:RQC}, henceforth named reality quantum correlator (RQC). A beta barium borate crystal (BBO) produces a pair of photons in the Bell state $\frac{1}{\sqrt{2}}(\ket{01}_{AB} + \ket{10}_{AB})$, with $\ket{01}_{AB}\equiv\ket{0}_A\ket{1}_B$ meaning that the polarization degrees of freedom $A$ and $B$ of the photons $\cal A$ and $\cal B$ are horizontal and vertical, respectively, and similarly for $\ket{10}_{AB}$.
Here, however, in order to analyze the role of the initial entanglement between $A$ and $B$, we consider the partially entangled state $\ket{\Psi_+}_{AB} =\mf{c}\ket{01}_{AB}
+\mf{s}\ket{10}_{AB}$ produced by the BBO plus an auxiliary equipment (BBO$'$) \cite{James2001}, with $\mf{c}=\cos(\theta/2)$, $\mf{s}=\sin(\theta/2)$, and $\theta\in\big[0,\frac{\pi}{2}\big]$. The photons are sent to spacelike separated regions, thus ensuring that they cannot causally influence each other during Alice's procedures. The photon $\cal B$ goes to a Mach-Zehnder interferometer (MZI) while the photon $\cal A$ moves toward a polarizing beam splitter (PBS$_{A}$), before which Alice can choose to include or not a quarter-wave plate (QWP). Being responsible for transforming the linear into circular polarization, this device will define the configurations referred to as QWP$_\text{in}$ or QWP$_\text{out}$ depending on whether it is placed in or out of photon $\cal A$'s path. The polarizing beam splitter (PBS) in photon $\cal B$'s path fully transmits (resp. reflects) horizontal (resp. vertical) polarization.
After that, a half-wave plate (HWP) disentangles the polarization $B$ from the other degrees of freedom. Next, $\cal{B}$ is submitted to nondestructive interactions with two other systems, here generically named atoms 1 and 2.  Some practical examples of how to encode a photon's which-way information without destroying it can be found in the literature \cite{Maffei2023, Distante2021, Huang2008, Lamas2001}. Prepared in their excited energy states $\ket{1}_{\epsilon_1}$ and $\ket{1}_{\epsilon_2}$, the atoms 1 and 2 are respectively inserted in the upper and lower paths of the MZI. Upon local interaction with the photon $\cal B$, one of the atoms will be stimulated to jump to the ground state ($\ket{0}_{\epsilon_{i}}$). Reflections take place in the mirrors (M) and the paths are recombined at the beam splitter (BS). The path degree of freedom of the photon $\cal A$ ($\cal B$),  hereafter referred to as $a$ ($b$), is measured by the detectors D$_{j,a}$ (D$_{j,b}$), with $j\in\{0,1\}$. The notation is such that a click in D$_{j,a}$ means that the $j$-th spatial mode of the photon $\cal A$ was $\ket{j}_a$, where $0$ ($1$) stands for the horizontal (vertical) spatial mode. As we show next, when Alice chooses the QWP$_\text{out}$ configuration and performs projective measurements with post-selection on $a$, the overall picture in Bob's site is such that the photon $\cal{B}$'s behavior becomes particle-like ($\mf{I}_b=0$), the atoms end up in a separable state, and the atoms' energies $\epsilon_{k}$ ($k\in\{1,2\}$) become elements of reality ($\mf{I}_{\epsilon_k}=0$). On the other hand, opting for  QWP$_\text{in}$ and performing again projective measurements with post-selection on $a$, Alice establishes maximum coherence to $b$, which leads this degree of freedom to maximally violate realism (i.e., $\mf{I}_b=1$) and the photon $\cal{B}$ to exhibit wave-like behavior~\cite{Dieguez2022}.
After passing the MZI and being detected, the photon $\cal{B}$ leaves the atoms 1 and 2 entangled, thus implying that $\epsilon_{k}$ are no longer elements of the physical reality ($\mf{I}_{\epsilon_k}=1$).

A related experiment was conducted previously \cite{Ma2013}, but a key distinction in our configuration is that the photon $\cal{B}$ interacts with atoms 1 and 2 inside the MZI. Reference~\cite{Araujo2024} can be viewed as an experimental realization of part of the present setup and also closing the loophole of locality. The authors conducted an optical implementation and artificially simulated the atoms by employing two extra degrees of freedom with the aid of beam displacers. In their experiment, they utilized the entanglement of three degrees of freedom of the same photon within the interferometer. Conversely, our approach includes the quanton traversing the second beam splitter (BS) and subsequent post-selection in the path domain, which guarantees that only the atoms stay entangled. Furthermore, the novel configuration described in this paper offers a process similar to the quantum teleportation protocol \cite{Bennett1993}, enabling Alice and Bob to generate a specific entangled state of the atoms, which can be utilized as a resource for quantum information processing and communication protocols. Moreover, in Ref.~\cite{Araujo2024}, the authors demonstrated that the entanglement generated between these degrees of freedom can act as an alternative measure for evaluating wave behavior within the interferometer, since only a delocalized system (wave) is capable of producing entanglement between far apart separated systems. This approach was highlighted in Refs.~\cite{Horodecki2009,Angelo2015} and was recently utilized to explore the quantumness of spacetime as seen in Refs.~\cite{Marletto2017,Bose2017}. This strategy implies that interferometric visibility is not the sole method for this purpose and is also not valid in general, as explored in Ref.~\cite{Starke2023}.
In addition, using interferometric visibility as a means of retrodiction measure has sparked controversies, including Wheeler's delayed choice \cite{Wheeler1984}, which suggests the concept of retrocausality, and Afshar's experiment~\cite{Afshar2007} that purported violation of Bohr's complementarity principle, allowing retroinference to resort to particle behavior. The entanglement of the atoms, as a witness of wave behavior, also addresses the unresolved question in Ref.~\cite{Starke2023_2} regarding the quantification of wave behavior within the interferometer throughout the framework of the generalized-entangled quantum eraser.

\section{Analysis and predictions}
\label{sec:analysis}

Moving forward with the formal analysis of the RQC, the whole system starts in $\ket{\Psi_0} =\ket{\Psi_+}_{AB}\ket{00}_{ab}\ket{ 11}_{\epsilon_1 \epsilon_2}$. 
Once the photon $\cal{B}$ enters the MZI and passes through the PBS, the state becomes
$
\ket{\Psi_1}=(i\mf{c}\ket{011}_{ABb}+\mf{s}\ket{100}_{ABb})\ket{011}_{a\epsilon_1 \epsilon_2} 
$
, with the PBS introducing a phase $e^{i \frac{\pi}{2}} = i$  \cite{Degiorgio1980, Zeilinger1981} in the reflected component, the one related to the vertical polarization $\ket{1}_B$. Then $\cal{B}$ passes through the HWP, which reverses the polarization and thus transforms $\ket{\Psi_1}$ into
\begin{equation}
\ket{\Psi_2} =\left(  i\mf{c}\ket{01}_{Ab}+\mf{s}\ket{10}_{Ab}\right)\ket{00}_{Ba}\ket{11}_{\epsilon_1 \epsilon_2} \label{eq:psi2}.
\end{equation}
One sees that the HWP disentangles $B$ from the other degrees of freedom, leaving only $A$ and $b$ entangled. That is, the entanglement initially encoded in the photons' polarizations (sector $AB$)  migrated to the sector $Ab$. At the next stage of photon $\cal{B}$'s route, the photon-atoms interaction (PAI) yields
$
\ket{\Psi_3}=\big( i\mf{c}\ket{01}_{Ab}\ket{ 10}_{\epsilon_1 \epsilon_2} + \mf{s}\ket{10}_{Ab}\ket{01}_{\epsilon_1 \epsilon_2}\big)\ket{00}_{Ba},
$ 
which indicates the formation of four-partite entanglement in the sector $Ab\epsilon_1\epsilon_2$.
Mirrors change the spatial mode with addition of a negligible global phase $e^{i \frac{\pi}{2}}$. Their actions transform the state $\ket{\Psi_3}$ into
$
\ket{\Psi_4}=\big(-\mf{c}\ket{00}_{Ab}\ket{10}_{\epsilon_1 \epsilon_2} +i\mf{s}\ket{11}_{Ab}\ket{01}_{\epsilon_1 \epsilon_2}\big)\ket{00}_{Ba}.
$ 
Since the BS's action leads $\ket{0}_b$ and $\ket{1}_b$ to 
$\ket{\omega_+}_b\equiv \frac{1}{\sqrt{2}}\left(\ket{0}_b+i\ket{1}_b \right)$ and 
$i\ket{\omega_-}_b\equiv \frac{1}{\sqrt{2}}\left(\ket{1}_b+i\ket{0}_b\right)$, respectively, the state becomes
\begin{equation}
\ket{\Psi_5}=\left(\mf{c}\ket{010}_{A\epsilon_1 \epsilon_2}\ket{\omega_+}_b +\mf{s}\ket{101}_{A \epsilon_1 \epsilon_2}\ket{\omega_-}_b\right)\ket{00}_{Ba}.\label{eq:psi5}
\end{equation}
While four-partite entanglement is maintained in the $Ab\epsilon_1\epsilon_2$ sector at the output of the MZI, quantum coherence now manifests itself in the $b$ eigenbasis.
Encompassing all the unitary dynamics of quantumness in our setup, equations~\eqref{eq:psi2} and \eqref{eq:psi5} will be the starting point for the analysis that follows, which consists of assessing how Alice's choice of QWP$_{\text{in}}$ or QWP$_{\text{out}}$ configuration correlates with the realism of observables accessible to Bob.
Before we continue, though, it is worth noting that entanglement with Alice's degrees of freedom renders the state accessible to Bob devoid of quantum correlations and coherences in the $\epsilon_{1,2}$ degrees of freedom. As a consequence, $\epsilon_k$ remain elements of reality since preparation, that is, $\mf{I}_{\epsilon_k}=0$ for all times. We now analyze whether and how this scenario changes upon Alice's choices.

Consider the QWP$_{\text{out}}$ configuration. When the photon $\cal{A}$ passes through PBS$_A$, $\ket{\Psi_2}$ transforms to
$\ket{\Psi_{2}^{\text{\tiny out}}} =i\big(\mf{c}\ket{001}_{Aab} +\mf{s}\ket{110}_{Aab}\big)\ket{011}_{B\epsilon_1\epsilon_2}.$ Every time detector D$_{0,a}$ clicks, the collapsed state reads $\ket{000111}_{AaBb\epsilon_1\epsilon_2}$. In these cases, the state accessible from Bob's location is $\Omega_2^\text{\tiny out}=\ket{\Omega_2^\text{\tiny out}}\!\bra{\Omega_2^\text{\tiny out}}$, where $\ket{\Omega_2^\text{\tiny out}}=\ket{0111}_{Bb\epsilon_1\epsilon_2}$. The dephasing map $\Phi_b$ has no effect in this case, that is, $\Phi_b(\Omega_2^\text{\tiny out})=\Omega_2^\text{\tiny out}$, which leads to $\mf{I}_b(\Omega_2^\text{\tiny out})=0$. Thus, upon the post-selection of $\ket{0}_a$, it follows that the path $b$ is an element of reality, which implies that Bob will not be able to verify wave-like behavior for $\cal{B}$ afterwards. In fact, it is straightforward to show that the state at the output of the MZI is $\ket{000\omega_+10}_{AaBb\epsilon_1\epsilon_2}$. Because the energy states do not have coherence and are not entangled, $\epsilon_{k}$ are elements of reality. All this demonstrates that $\cal{B}$  traveled a definite arm of the MZI and did not couple with atom 1 (particle-like behavior).

\begin{figure*}[ht]
    \centering
    
    \includegraphics[scale=1.00]{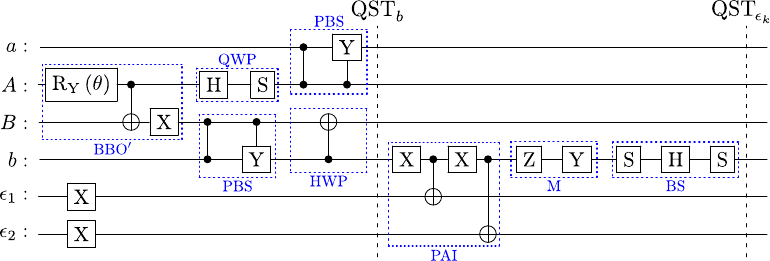}
    \caption{The quantum circuit implemented in the IBMQ for simulating the reality quantum correlator (RQC).  Each blue dotted box is the decomposition into quantum logic gates of the optical devices and the photon-atoms interaction (PAI) explored in the main text. Each horizontal line represents a qubit that is identified with the notation used in the main text, and all qubits are initialized in the state $\ket{0}$. 
    The photons $\mathcal{A}$ and $\mathcal{B}$ possess respective path degrees of freedom, $a$ and $b$, and the polarization degrees of freedom, $A$ and $ B$.
    The barium beta borate plus auxiliary equipment (BBO$^{\prime}$), which is simulated in quantum computers through the action of the RY$(\theta)$ gate on $A$, the CNOT gate having $A$ as control and target in $B$, and the Pauli $X$ gate acting on $B$.
    After that, the entangled state in the polarization degrees of freedom is given by $\ket{\Psi_+}_{AB} =\mf{c}\ket{01}_{AB}
    +\mf{s}\ket{10}_{AB}$ with $\mf{c}=\cos(\theta/2)$, $\mf{s}=\sin(\theta/2)$, and $\theta\in\big[0,\frac{\pi}{2}\big]$.
    The photon $\mathcal{B}$ enters the Mach-Zehnder interferometer (MZI) passing first through the polarizing beam splitter (PBS) which, as PBS$_{A}$, is implemented with the gates $C_Z$ and $C_Y$, both with polarization as control and the path as target. Then $\mathcal{B}$ passes through the half-wave plate (HWP), which can be implemented through a CNOT with control in $b$ and target in $B$. The gate $X$ is applied to the qubits $\epsilon_1$ and $\epsilon_2$ to produce the initial excited state of the energy of the atoms $\ket{11}$ considered in our proposal. The photon-atoms interaction (PAI) is constructed using CNOTs, both with control in $b$ and target in $\epsilon_1$ or $\epsilon_2$ (for $\epsilon_1$, though, the gate $X$ has to be applied before and after control in the CNOT, as we want the interaction to occur when the path $b$ is in state $\ket{0}$, corresponding to the upper path of the MZI). The mirrors' (M) combined action is implemented using $U_\text{M}^b = YZ$ and the beam splitter (BS) is simulated with $U_{\text{BS}}^b = SHS$. Finally, the quarter-wave plate (QWP) is implemented via $U_{\text{QWP}}^A = SH$, the device that represents Alice's choice, which can be present (QWP$_{\text{in}}$) or absent (QWP$_{\text{out}}$) in the experimental setup. The dashed vertical lines represent quantum state tomography (QST). The first run, referred to as QST$_b$, was performed when the system is in the state $\ket{\Psi_2^\text{\tiny out(in)}}$.
    The second run of QST, denoted QST$_{\epsilon_{k}}$,
    was performed when the state of the system is $\ket{\Psi_5^\text{\tiny out(in)}}$.
    }
\label{fig:qcircuit}
\end{figure*}

In the QWP$_\text{in}$ configuration, QWP converts $\ket{0}_A$ and $\ket{1}_A$ to $\frac{1}{\sqrt{2}}(\ket{0}_A+i\ket{1}_A)$ and $\frac{1}{\sqrt{2}}(\ket{0}_A-i\ket{1}_A)$, respectively. As a consequence, the state in Eq.~(\ref{eq:psi2}), transforms into
$\ket{\psi_{2}^\text{\tiny in}}=\frac{1}{\sqrt{2}}\big(\ket{0}_{A}\ket{\beta_+}_b-i\ket{1}_{A}\ket{\beta_-}_b\big)\ket{0011}_{aB\epsilon_1\epsilon_2}
$, where $\ket{\beta_\pm}_b\equiv\mf{s}\left\vert 0\right\rangle _{b}\pm i\mf{c}\left\vert 1\right\rangle _{b}.$ Finally, after the PBS$_A$, the state of the system results in $\ket{\Psi_2^\text{\tiny in}}=\frac{1}{\sqrt{2}}\big(\ket{00}_{Aa}\ket{\beta_+}_b + \ket{11}_{Aa}\ket{\beta_-}_b\big)\ket{011}_{B\epsilon_1\epsilon_2}$. Now, every time D$_{0,a}$ clicks, the state reduces to $\ket{000\beta_+11}_{AaBb\epsilon_1\epsilon_2}$, implying that the part accessible to Bob is $\Omega_2^\text{\tiny in}=\ket{\Omega_2^\text{\tiny in}}\!\bra{\Omega_2^\text{\tiny in}}$, with $\ket{\Omega_2^\text{\tiny in}}=\ket{0\beta_+11}_{Bb\epsilon_1\epsilon_2}$. The fundamental difference from the previous case is now apparent: here, one has $\Phi_b(\Omega_2^\text{\tiny in})\neq \Omega_2^\text{\tiny in}$ because the dephasing map $\Phi_b$ destroys the coherence of the paths in $\ket{\beta_+}_b$, so one finds $\mf{I}_b(\Omega_2^\text{\tiny in})=-\mf{c}^{2}\log_{2}\mf{c}^{2}-\mf{s}^{2}\log_{2}\mf{s}^{2}$.
Clearly, the irreality equals the amount of entanglement encoded in the initial state $\ket{\Psi_+}_{AB}$ (as quantified by the von Neumann entropy of the reduced state).
This means that maximum entanglement $\big(\mf{c} = \mf{s} =1/\sqrt{2}\big)$ implies that there will be no element of reality for photon $\cal{B}$'s path. Note that $b$ irrealism is here ensured solely by quantum coherence, since the $b$ partition is uncorrelated with the others. To verify this, Bob needs to certify entanglement between $\epsilon_1$ and $\epsilon_2$. Direct calculations show that the state at the output of the MZI reads $\frac{1}{\sqrt{2}}\ket{000}_{AaB}\big(\ket{1}_b\ket{\xi_-}_{\epsilon_1\epsilon_2}+i\ket{0}_b\ket{\xi_+}_{\epsilon_1\epsilon_2} \big)$, where we define $\ket{\xi_\pm} _{\epsilon_1 \epsilon_2}\equiv \mf{s}\left\vert 01\right\rangle_{\epsilon_1 \epsilon_2}\pm \mf{c}\left\vert 10\right\rangle _{\epsilon_1 \epsilon_2}$. Hence, via post-selection on, say $\ket{0}_b$, Bob can verify the amount $-\mf{c}^{2}\log_{2}\mf{c}^{2}-\mf{s}^{2}\log_{2}\mf{s}^{2}$ of entanglement in $\ket{\xi_+}_{\epsilon_1\epsilon_2}$ and conclude that $\epsilon_{k}$ are not elements of reality. The breakdown in reality is here induced solely by quantum discord, which is known to be equal entanglement for a pure state like $\ket{\xi_+}_{\epsilon_1\epsilon_2}$. It is worth noting that had we opted to use only one atom, we would have a simpler version of the RQC in which Bob would be able to validate the wave-like behavior of $\cal{B}$' through a different quantum resource. Indeed, it is not difficult to show that in such a case Bob would find quantum coherence in the atomic energy.

\begin{table*}[th]
    \def\S{2cm}
    \begin{tabular}{lp{\S}p{\S}p{\S}p{\S}p{\S}p{\S}p{\S}}
        \hline\hline
        ibm\_nairobi parameters & Q0 & Q1 & Q2 & Q3 & Q4 & Q5 & Q6 \\
        \hline
        Frequency (GHz) & 5.26 & 5.17 & 5.274 & 5.027 & 5.177 & 5.293 & 5.129 \\
        T1 ($\mu$s) & 111.64 & 86.2 & 98.41 & 93.04 & 109.38 & 97.2 & 115.21 \\
        T2 ($\mu$s) & 29.42 & 58.73 & 107.66 & 10.4 & 70.71 & 20.16 & 108.19 \\
        Readout error ($10^{-2}$) & 2.500 & 5.970 & 2.330 & 2.170 & 3.150 & 2.500 & 2.770 \\
        Pauli-$X$ error ($10^{-4}$) & 3.040 & 5.095 & 4.801 & 6.167 & 2.422 & 3.607 & 1.942 \\
        CNOT error ($10^{-2}$) & 0\_1:\,1.114 & 1\_3:\,1.388 \newline 1\_2:\,1.508 \newline 1\_0:\,1.114 & 2\_1: 1.508 & 3\_5:\,1.88 \newline 3\_1:\,1.388 & 4\_5:\,0.514 & 5\_6:\,0.672 \newline 5\_4:\,0.514 \newline 5\_3:\,1.88 & 6\_5:\,0.672 \\
        Gate time (ns) & 0\_1:\,248.89 & 1\_3:\,270.22 \newline 1\_2:\,426.67 \newline 1\_0:\,284.44 & 2\_1:\,391.11 & 3\_5:\,277.33 \newline 3\_1:\,305.78 & 4\_5:\,312.89 & 5\_6:\,341.33 \newline 5\_4:\,277.33 \newline 5\_3:\,241.78 & 6\_5:\,305.78 \\
        \hline\hline
    \end{tabular}
    \caption{Calibration data for the IBMQ Falcon processor ibm\_nairobi quantum chip.}
    \label{tab_nairobi}
\end{table*}

So far, we have demonstrated that (i) the configuration selected by Alice is correlated with the realism of $b$, and (ii) Bob can confirm photon $\cal{B}$'s wavelike behavior by examining the quantum resources encoded in the atoms' energy states. Now we consider a scenario wherein Alice's intervention is temporarily postponed. We start with the state in Eq.~\eqref{eq:psi5} and first consider the QWP$_\text{out}$ configuration. By passage of $\cal{A}$ through the PBS$_A$, the state $\ket{\Psi_5}$ changes to $\ket{\Psi_5^\text{\tiny out}}=\big(\mf{c}\ket{0010}_{Aa\epsilon_1\epsilon_2}\ket{\omega_+}_b+i\mf{s}\ket{1101}_{Aa\epsilon_1\epsilon_2}\ket{\omega_-}_b \big)\ket{0}_B$. 
Measurements of $a$ and $b$ post-selected in $0$ result in $\ket{000010}_{AaBb\epsilon_1\epsilon_2}$.
In this case, the state accessible to Bob is $\Omega_5^\text{\tiny out}=\ket{\Omega_5^\text{\tiny out}}\!\bra{\Omega_5^\text{\tiny out}}$ with $\ket{\Omega_5^\text{\tiny out}}=\ket{0010}_{Bb\epsilon_1\epsilon_2}$.
The action of either $\Phi_{\epsilon_1}$ or $\Phi_{\epsilon_2}$ is innocuous, because the energies are neither entangled nor in superposition.
It immediately follows that $\mf{I}_{\epsilon_k}(\Omega_5^\text{\tiny out})=0$, meaning that $\epsilon_k$ are elements of reality. Considering now the QWP$_\text{in}$ configuration, after the system evolves to the state $\ket{\Psi_5}$, photon $\cal{A}$ passes through QWP and PBS$_{A}$. The resulting state is $\ket{ \Psi_{5}^{\text{\tiny in}}} =\frac{1}{2}\big(\ket{000}_{Aab}\ket{\xi_+}_{\epsilon_1 \epsilon_2} - i\ket{001}_{Aab}\ket{\xi_-}_{\epsilon_1 \epsilon_2} + \ket{110}_{Aab}\ket{\xi_-}_{\epsilon_1 \epsilon_2} - i\ket{111}_{Aab}\ket{\xi_+}_{\epsilon_1 \epsilon_2}\big)\ket{0}_B$. Assuming again projection and post-selection on $\ket{00}_{ab}$, we arrive at $\ket{0000}_{AaBb}\ket{\xi_+}_{\epsilon_1\epsilon_2}$. In this case, the state accessible from Bob's location is $\Omega_5^\text{\tiny in}=\ket{\Omega_5^\text{\tiny in}}\!\bra{\Omega_5^\text{\tiny in}}$ with $\ket{\Omega_5^\text{\tiny in}}=\ket{00}_{Bb}\ket{\xi_+}_{\epsilon_1\epsilon_2}$. Clearly, the energies are left entangled. As a consequence,  $\mf{I}_{\epsilon_k}(\Omega_5^\text{\tiny in})=-\mf{c}^{2}\log_{2}\mf{c}^{2}-\mf{s}^{2}\log_{2} \mf{s}^{2}$, implying that the energies will not be elements of reality in general. As already mentioned, the irrealities are numerically equal to the amount of entanglement encoded in both the initial state $\ket{\Psi_0}$ and $\ket{\xi_+}_{\epsilon_1\epsilon_2}$. Having arrived at the same conclusions as in the previous analysis, we demonstrated that the timing of Alice's actions plays no relevant role in our claim that they can correlate with the reality of Bob's observables, so no fine-tuning in time is experimentally required. Furthermore, notice that such effect does not allow for signaling, for Bob's certification can be done only after classically communicating with Alice.

\begin{figure}[ht]
    \centering
    \includegraphics[scale=0.5]{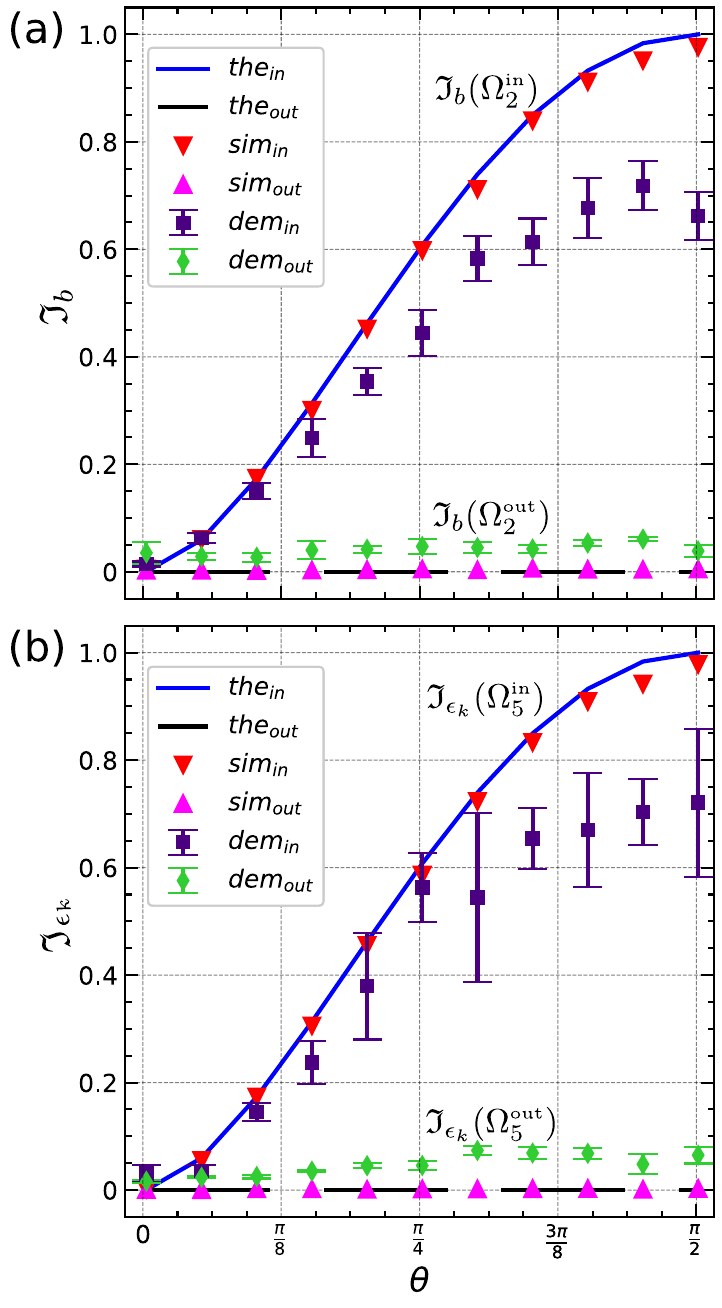}
    \caption{Theoretical (lines) and simulated and demonstrative (points) results for the irreality as a function of $\theta$, the parameter related to the initial polarization entanglement. The simulation results, obtained via classical emulation of the quantum circuit, are denoted by triangular points, whereas the experimental results are given by diamond and square points. (a) Irreality of the photon $\mathcal{B}$ path, $b$, calculated from quantum state tomography applied after the HWP and before $\mathcal{B}$ interacts with the atoms 1 and 2. (b) Irrealities of the atoms' energies ($\mf{I}_{\epsilon_{k}}$) calculated for the quantum state obtained at the MZI output. Both graphs present the results considering the scenarios QWP$_{\text{in}}$ and QWP$_{\text{out}}$. The simulated and experimental points around the solid theoretical line refer to the case where the QWP is inserted in the device (QWP$_{\text{in}}$). The points around the dashed theoretical line are for cases where QWP is not present in the experimental apparatus (QWP$_{\text{out}}$). For $\theta = \pi/2$ (resp. $\theta=0$) the initial entanglement between the polarizations $A$ and $B$ is maximum (resp. zero). The error bars are the standard deviation for ten repetitions of the demonstration. For the demonstrations using IBMQ, we employed the Falcon processor ibm\_nairobi quantum chip (its calibration parameters are given in Table~\ref{tab_nairobi}).}
\label{fig:irr}
\end{figure}

\section{EXPERIMENTAL RESULTS ON QUANTUM
COMPUTERS}
\label{sec:QC}

In what follows, we will demonstrate the proposed experiments using one of the IBM's quantum computers (IBMQ). The calibration data for the Falcon processor ibm\_nairobi quantum chip we used here are shown in Table~\ref{tab_nairobi}. In order to reproduce the RQC of Fig.~\ref{fig:RQC} we use six qubits; see Fig.~\ref{fig:qcircuit}. The main idea of implementing an optical device in a quantum computer is to obtain the unitary matrix of the optical device and convert it into quantum logic gates (further details about these implementations and associated quantum simulations are available in Refs.~\cite{Starke2023,Starke2023_2}). Constrained by the number of allowable executions for a given quantum circuit, we conducted quantum state tomography (QST) on a subset of qubits.
While all the setup simulations included all six qubits ($a,A,b,B,\epsilon_1$, and $\epsilon_2$) in the IBMQ as theoretically described, the QST was performed only on $A,b,\epsilon_1$, and $\epsilon_2$. This is perfectly admissible because $a$ is maximally correlated with $A$ (so measuring one gives the other) and $B$ is fully uncorrelated. Indeed, theoretically, one can verify that the irreality does not change with this simplification.

The QST was executed at two distinct stages of the circuit (indicated by vertical dashed lines in Fig.~\ref{fig:qcircuit}) for the QWP$_{\text{out}}$ and QWP$_{\text{in}}$ configurations.
The first run, referred to as QST$_b$, was performed when the system is in the state $\ket{\Psi_2^\text{\tiny out(in)}}$.
The subsequent run, labeled QST$_{\epsilon_{k}}$, took place at the MZI output, where the state of the system is $\ket{\Psi_5^\text{\tiny out(in)}}$.
In both cases, we performed the QST of $\mathcal{A}$ after the PBS$_{A}$, with $a$ and $B$ discarded.
In the first case, we calculated the irreality of $b$ and in the second case we obtained the irrealities of the atomic energies $\epsilon_1$ and $\epsilon_2$. The results are shown, respectively, in Figs.~\ref{fig:irr}~(a) and \ref{fig:irr}~(b), where the $\theta$ parameter controls the amount of initial entanglement between the polarization of $\mathcal{A}$ and $\mathcal{B}$.
It is well known that current \cite{IBMQ} quantum computers still have associated errors due to couplings with the environment, as well as thermal imperfections and imperfections of devices \cite{Papic2023}.
Because of this, we use Qiskit tools to mitigate \cite{McKay2018} the errors obtained at the demonstration points.
In particular, the substantial variation in the data obtained in Fig.~\ref{fig:irr}~(b) arises due to a greater depth of the quantum circuit, particularly for the CNOT gates used to simulate the PAI. Systems developed by IBM automatically integrate SWAP gates dependent on qubit interconnections to facilitate correlation gates between qubits. This often results in a further deepened quantum circuit, which can also ultimately be affected by the loss of the system's  coherence.
The demonstrative results are seen to agree fairly well with the theory and validate our claim that there is a correlation of the quantum reality of the observables in Bob's laboratory with causally disconnected choices made at Alice's laboratory. The results show that the elements of reality of the path and the atoms' energies are impacted by different quantum resources, namely quantum coherence and quantum discord, respectively. Moreover, the degree of irreality in each case was always numerically equal to the amount of entanglement encoded in the initial state.

\section{Final Remarks}
\label{sec:FR}

In summary, our work suggests that nature allows the occurrence of strong correlations between the local destruction of elements of reality and causally disconnected actions, in stark contrast to the EPR idea motivated by relativistic locality. This conclusion is supported by theoretical predictions of quantum mechanics, aligned with a well-established notion of quantum realism, and a proof of principle executed on IBM's quantum computers.
We hope that our results may motivate researchers to implement our experiment proposal in actual optical settings with individual subsystems as atoms and to close the locality loophole that we have in our IBMQ demonstration.
Methods for closing the ``freedom of choice'' loophole are also available~\cite{Ma2013}. In addition, it is worth noting that the degrees of freedom represented here by atomic energies could eventually be integrated into communication protocols, such as in quantum teleportation contexts or even in tasks involving the generation of bipartite entanglement without direct interaction.

\begin{acknowledgments}
\textit{Acknowledgments:} We thank Luiz C. Ryff, Alexandre D. Ribeiro, Felipe E. L. da Cruz, and Douglas F. Pinto for helpful discussions.

\textit{Funding:} R.M.A. and J.M. acknowledge support of the
National Institute for the Science and Technology of Quantum Information (INCT-IQ),
Grant No. 465469/2014-0.
R.M.A. was supported by the
National Council for Scientific and Technological Development (CNPq),
Grant No. 305957/2023-6.
J.M. was supported by CNPq,
Grant No. 309862/2021-3.
D.S.S. was supported by the
Coordination for the Improvement of Higher Education Personnel (CAPES),
Grant No. 88887.827989/2023-00.
The authors acknowledge support from CNPq through
Grant No. 409673/2022-6. 
\end{acknowledgments}


\end{document}